\documentstyle[multicol,prl,aps,psfig]{revtex}

\begin{document}

\title{The Adsorption of Atomic Nitrogen on Ru\,(0001):
Geometry and Energetics}

\author{
S. Schwegmann, A. P. Seitsonen, H. Dietrich, H. Bludau,
H. Over\cite{Herbert}, K. Jacobi, and G. Ertl}
\address{Fritz-Haber-Institut der Max-Planck-Gesellschaft,
Faradayweg 4-6, D-14195 Berlin, Germany}

\maketitle

\begin{abstract}
   The local adsorption geometries of the ($2\times 2$)--N and the
($\sqrt{3}\times\sqrt{3})$R30$^\circ$--N phases on the Ru\,(0001) surface
are determined by analyzing low-energy electron diffraction (LEED)
intensity data. For both phases, nitrogen occupies the threefold hcp
site. The nitrogen sinks deeply into the top Ru layer resulting in a
N--Ru interlayer distance of 1.05~\AA{} and 1.10~\AA{} in the
($2\times 2$) and the ($\sqrt{3}\times \sqrt{3}$)R30$^\circ$ unit cell,
respectively. This result is attributed to a strong N binding to the
Ru surface (Ru--N bond length = 1.93~\AA{}) in both phases as
also evidenced by ab-initio calculations which revealed binding
energies of 5.82~eV and 5.59~eV, respectively.
\end{abstract}

\vspace*{2mm}

\noindent Submitted to Chem. Phys. Lett.

\vspace*{2mm}

\begin{multicols}{2}

  The rate-limiting step in the Haber-Bosch synthesis (ammonia
synthesis) is represented by the dissociation process of molecular
nitrogen over the active metal surfaces [1]. While for the Fe(111)
surface the dissociative sticking coefficient of nitrogen is about
10$^{-6}$ [2], the Ru surfaces exhibit only an exceedingly small
sticking coefficient of about 10$^{-12}$ at room temperature [3]. Yet,
the promoted Ru catalyst is even more efficient than the industrially
used Fe catalyst [4] since blocking of the surface by too much
nitrogen and/or poisoning by traces of oxygen is less severe. Apart
from the low sticking probability, the turn over rates of the ammonia
reaction are determined by the energy barriers of adsorbed hydrogen
and nitrogen to form NH$_x$. This barrier is governed by the
adsorption energy and the adsorption geometry of atomic nitrogen since
N is much more strongly bound to Ru than H. In this Letter, we present
the atomic geometry and energetics of the atomic nitrogen adsorbed on
Ru\,(0001) at two distinct coverages by employing the techniques of
low-energy electron diffraction (LEED) and density-functional theory
(DFT).

  The experiments were conducted in a UHV system (base pressure
$2.0\times 10^{-10}$~mbar) equipped with a 4-grid LEED optics, a CMA
for Auger spectroscopy and a quadrupole mass spectrometer for thermal
desorption spectroscopy (TDS). The Ru\,(0001) crystal could be heated
resistively up to 1500~K and cooled to 130~K using liquid nitrogen;
the temperature was measured with a NiCr--Ni thermocouple spot welded
to the back of the sample. Further details about the experimental
setup can be found elsewhere [7]. The Ru\,(0001) crystal was cleaned in
a standard way by Ar$^+$ sputtering and repeated cycles of oxygen
exposure followed by annealing to 1560~K in order to remove carbon
contamination. Since the dissociative sticking coefficient of
molecular nitrogen is extremely small we used ammonia and its ability
to decompose at the Ru\,(0001) surface above 350~K into atomic nitrogen
and hydrogen which desorbs at the chosen sample temperature.
Nevertheless, NH$_3$ exposures of several thousands of Langmuirs were
needed to accommodate substantial amounts of atomic nitrogen at the Ru
surface.

  More specifically, the atomic nitrogen overlayers were prepared by
adsorbing ammonia, which is known to dissociate into N and NH already
during chemisorption above the desorption temperature of NH$_3$. There
is a delicate balance for the uptake of atomic nitrogen: NH$_3$ dosing
at lower temperatures results in a higher saturation coverage of
atomic nitrogen but at the expense of very high NH$_3$ doses, while at
higher adsorption temperature the uptake of atomic nitrogen takes
place rapidly reaching a smaller saturation coverage. To get the
required high N coverage on a reasonable time scale, the temperature
was lowered step by step starting from 580~K to room temperature while
dosing a total amount of about 20000~L of NH$_3$ by backfilling the
UHV chamber with $1.0\times 10^{-5}$~mbar NH$_3$. The specific protocol of
NH$_3$ dosing can be found in Ref. [8]. After this procedure it took
typically 12 hours to retain a base pressure of $2.0\times
10^{-9}$~mbar (the main constituent of the residual gas was still due
to NH$_3$) as ammonia easily sticks to and desorbs from the walls of
the UHV chamber.

  The ($\sqrt{3}\times\sqrt{3}$)R30$^\circ$ and ($2\times 2$) phases
of nitrogen were prepared by dosing NH$_3$ as described above and by
heating the sample to 525~K and 615~K for the
($\sqrt{3}\times\sqrt{3}$)R30$^\circ$ and the ($2\times 2$) phase,
respectively (heating rate 5~K/s). From HREELS measurements it is
known that for temperatures beyond 500~K neither NH nor NH$_2$ species
are left on the surface [8].

  After cooling the sample to 140~K LEED exhibited well-ordered
($2\times 2$) and ($\sqrt{3}\times\sqrt{3}$)R30$^\circ$ patterns.
LEED~I(V) curves were taken at normal incidence. Four fractional-order
beams and two integer-order beams in an energy range from 50~eV to
300~eV were collected in the case of the ($2\times 2$), three
fractional-order beams and two integer-order beams in an energy range
from 30~eV to 350~eV for the ($\sqrt{3}\times\sqrt{3}$)R30$^\circ$.
The measurements were recorded on-line from the fluorescent screen of
a display type 4-grid LEED optics by using a computer-controlled
video-LEED system [9]. The experimental data were further processed by
averaging over symmetrically equivalent beams, smoothing, and
correcting for constant emission current.

  Due to re-adsorption of NH$_3$ from the residual gas it was
necessary to heat the sample again to 525~K
[($\sqrt{3}\times\sqrt{3}$)R30$^\circ$-N] or 615~K [($2\times 2$)-N]
after a cooling and measuring period of about 5 minutes. This
precaution ensures that the NH$_3$-induced changes in the LEED~I(V)
curves of fractional-order beams remained small as quantified by a
Pendry r-factor [10] of R$_{\rm P}$~$<$~0.10; note that the
experimental uncertainty due to re-preparation was found to be about
R$_{\rm P} \approx 0.08$.

  To evaluate the experimental I(V) data, fully-dynamical LEED
calculations were performed using the program package of Moritz [11].
Nine spin-averaged, relativistically corrected phase shifts were used
for the Ru substrate [12], and the phase shifts of nitrogen were taken
from the literature [13] which have been already successfully applied
to the analysis of molecular nitrogen on the clean and O-precovered
Ru\,(0001) surface [14]. The influence of thermal vibrations on the LEED
intensities was accounted for by correcting the phase shifts with a
fixed effective Debye temperature of 400~K for the substrate and for
nitrogen.

  The agreement between experimental and calculated I(V) curves was
quantified by the RDE factor introduced by Kleinle {\it et al.}\/ [15]
and by Pendry's r-factor R$_{\rm P}$ [10]. The refinement of
structural parameters and of the real part of the muffin-tin zero was
carried out by employing a nonlinear least-squares optimization scheme
[16] based on the least-squares sum either of the LEED intensities or
of the corresponding Y~functions.

  For the density-functional theory (DFT) calculations we employed the
generalized gradient approximation (GGA) of Perdew et al.\/ [17] for
the exchange-correlation functional and used ab-initio
pseudo-potentials created by the scheme of Troullier and Martins in
the fully separable form [18]; the electronic wave functions were
expanded in a plane-wave basis set. The surface was modeled using the
supercell approach, incorporating five layers of Ru\,(0001) where the N
atoms were adsorbed on one side of this slab [19]. Calculations were
performed using a ($2\times 2$), a
($\sqrt{3}\times\sqrt{3}$)R30$^\circ$ and a (1$\times$1) surface unit
cell with an energy cut-off of 50~Ry. The integral over the Brillouin
zone was performed using a special k-point set [20], and care was
taken to ensure an equivalent sampling in all (surface) geometries,
with 9 k-points in the irreducible part of the (1$\times$1) Brillouin
zone. Fermi broadening of the occupation numbers was done with a width
of 0.1~eV. The calculation scheme allows for simultaneous relaxation
of the electrons and atoms, where we relaxed the positions of the N
atoms and the atoms in the top Ru layer, keeping the lower three Ru
layer spacings fixed at the bulk values.

  To estimate the accuracy of the calculations, we studied the clean
substrate Ru\,(0001) and the free N$_2$ molecule. The lattice constant
of Ru obtained turned out to be 2.78~\AA{}, slightly larger than in
experiments (2.71~\AA{}) as usually observed with GGA. Therefore the c/a
ratio was somewhat smaller than the experimental value, being 1.5755.
When relaxing the (0001) slab, the outermost layer spacing contracts
by about 3~\%, a value which is also slightly larger than the
experimental value of 2$\pm$1~\% [21]. For the very strong bond of the
N$_2$ molecule we obtained a binding energy of 10.24~eV (not corrected
for zero-point motion), a N--N bond length of 1.104~\AA{}, and a N--N
vibrational frequency of 2336~cm$^{-1}$. These values compare well
with the experiment of values of: 9.77~eV, 1.097~\AA{}, and
2359~cm$^{-1}$.

  The LEED analyses were performed by placing the nitrogen in both
phases in high-symmetry sites -- i.e., bridge, on-top, threefold
hollow fcc and hcp -- and varying the N--Ru layer spacing, the topmost
Ru--Ru layer distance as well as the lateral and perpendicular
displacements in the top Ru layer as long as they are compatible with
the symmetry of the unit cell. To explore a wide range of parameter
space, we changed, in a first step, the principal structure parameter,
{\it i.\,e.}, the N--Ru layer spacing, from 0.5~\AA{} to 2.5~\AA{} in
steps of 0.1~\AA{}, keeping the other co-ordinates of the Ru atoms at
those values found for the clean Ru\,(0001) surface. Starting from the
optimum value of the N--Ru layer distance, we then performed an
automated refinement varying the structural parameters mentioned above
and the real part of the muffin-tin potential. The optimum r-factors
reached for each of these high-symmetry configurations are compiled in
Table~I, from which it is obvious that in both phases the hcp site
represents the actual adsorption site. This result is consistent with
a recent STM study in which the adsorption site of nitrogen was
determined by labeling hcp-sites with coadsorbed oxygen atoms [22].

  The structural parameters of the best-fit models are presented in
Figs.~1 and 2, and the good agreement between calculated and
experimental LEED IV data is illustrated in Figs.~3 and 4. The
chemisorption of nitrogen induces small lateral displacements by
0.06 $\pm$ 0.05~\AA{} ($2\times 2$) and 0.02 $\pm$ 0.05~\AA{}
($\sqrt{3}\times\sqrt{3}$)R30$^\circ$ in the topmost Ru layer which are
small if compared to corresponding displacements of about 0.1~\AA{}
due to the chemisorption of oxygen on Ru\,(0001) [23]. The N--Ru layer
spacing turned out to be 1.05~\AA{} and 1.10~\AA{} in the ($2\times
2$) and the ($\sqrt{3}\times\sqrt{3}$)R30$^\circ$ phases,
respectively. These values are substantially smaller than those found
with oxygen phases (about 1.24~\AA{}), which is attributed to a
stronger bonding of N to Ru. In TDS, however, nitrogen comes off the
surface in the temperature range from 500~K to 850~K, while oxygen
desorbs above 1000~K. Yet, both experimental observations are not
conflicting since the heat of formation of molecular nitrogen
(9.77~eV) is much higher than that of molecular oxygen (5.16~eV). Note
that TDS yields only information about the activation energy of
desorption which, to a first approximation, is the difference of the
Ru--X binding energy and one half of the binding energy of X$_2$ (X
either N or O). Ab-initio calculations presented here support this
view and provide the Ru--N binding energies in the (2$\times$2)-N,
($\sqrt{3}\times\sqrt{3}$)R30$^\circ$-N, and (1$\times$1)-N phases to
be 5.82~eV, 5.59~eV, and 4.90~eV, respectively, which are indeed
higher than those found in the ($2\times 2$)-O, (2$\times$1)-O, and
(1$\times$1)-O phases (5.55~eV, 5.28~eV, and 4.80~eV) [24]. In all
these cases, DFT calculations showed that adsorption takes place in
the hcp site. Note that the (1x1)-N phase is not thermodynamically
stable as discussed below.

  In contrast to the ($2\times 2$)-O and (2$\times$1)-O phases on
Ru\,(0001), which show a slight contraction of the top Ru-Ru layer
spacing, the averaged top Ru--Ru interlayer spacing in the
($\sqrt{3}\times\sqrt{3}$)R30$^\circ$-N phase is slightly expanded by
1~\% indicating that moderate amounts of nitrogen are sufficient to
deplete the population of bonding d orbitals between the first and
second Ru layer, thus removing the contraction of the clean Ru\,(0001)
surface. The hard-sphere radii of N found on the Ru\,(0001) surface
are 0.58~\AA{} in both phases which compares well with values already
found on other metal surfaces, such as Ni\,(110)-(2$\times$3)-N
(0.59~\AA{}) [25], Cu\,(110)-(2$\times$3)-N (0.62~\AA{}) [26], and
Rh\,(110)-(2$\times$1)-N (0.57~\AA{}) [27].

  In Table~II, the main structural parameters of both nitrogen phases
obtained by ab-initio calculations are compared to those obtained by
LEED. Within the quoted error bars these parameters are identical.
Even details of the N structures are reconciled with DFT calculations
such as the atomic displacements of the Ru atoms in the ($2\times
2$)-N phase: While LEED revealed a rumpling of 0.07 $\pm$ 0.04~\AA{}
and a lateral shift of 0.06 $\pm$ 0.05~\AA{}, DFT calculations
determined these displacements to be 0.086~\AA{} (vertical) and
0.067~\AA{} (lateral). In addition to these experimentally observed N
surface structures, DFT calculations for a hypothetical (1$\times$1)-N
structure were carried out. The Ru--N layer spacing (1.24~\AA{})
turned out to be larger than that for the ($2\times 2$)-N and the
($\sqrt{3}\times\sqrt{3}$)R30$^\circ$-N phases, and the Ru--Ru layer
spacing is heavily expanded by 5.2~\% with respect to the bulk value.
This investigation was triggered by a recent DFT/LEED investigation of
the (1$\times$1)-O phase on Ru\,(0001) [24]. Also with the
(1$\times$1)-O structure it was found that the first Ru--Ru layer
distance is substantially expanded by 3.5~\%. The main difference
between the (1$\times$1)-N and the (1$\times$1)-O phases on Ru\,(0001)
is, however, that according to DFT calculations, the (1$\times$1)-O is
exothermic with respect to molecular oxygen, while the (1$\times$1)-N
phase is not. The binding energy of nitrogen in the (1$\times$1) phase
is only 4.52~eV which is by 0.6~eV smaller than half the formation
energy of molecular nitrogen.

  To determine the critical N-coverage which can be thermally
stabilized on the Ru\,(0001) surface, we performed DFT calculations of
the (2x1)-N and the (2x2)-3N overlayers assuming hcp-adsorption only.
It turned out that nitrogen in the (2x1)-N is stabilized by 0.2~eV
with respect to half the formation energy of N$_2$, while nitrogen in the
(2x2)-3N phase is endothermic by 0.2~eV. This result suggests the
nitrogen saturation coverage to be about 0.5 (so far not observed in
experiments) which is also compatible with the formation of heavy wall
domain boundaries as proposed for N overlayers with coverages larger
than 0.33 [8].

  The strong bonding of nitrogen to Ru\,(0001) [exemplified for the
(2x2)-N overlayer] can be explained most easily by using a simple
two-level tight-binding model [28]. The same model has recently shown
to work for O/Rh\,(110) [29] and O/Ru\,(0001) [30]. In this model the
2p orbitals of N interact with d-orbitals of Ru close to the Fermi
energy (cf. Fig. 5), forming a bonding hybride at an energy slightly
lower than of the 2p orbital of the free N atom and an anti-bonding
hybride whose energy is 3~eV above the Fermi energy. The bonding
hybride consists (mainly) of sp hybride orbitals of the N atom and
d$_z{^2}$-like orbitals of the three nearest-neighbor Ru atoms,
aligned along the bond axis between the N and Ru atoms. In the
non-bonding hybride, at energies $-$3.5~eV $<\ \varepsilon\ <$ $-$2.5~eV
below Fermi, there are one-electron states with pz-character at the N
atom and d character at the Ru atoms. Close to the Fermi energy many
d$_z{^2}$-like orbitals have been emptied upon forming the
anti-bonding hybride, at 3~eV above Fermi. The corresponding wave
functions exhibit a distinct node between the N atoms and the
nearest-neighbor Ru atoms thus indeed demonstrating anti-bonding
character.

  The strong bond of N on Ru\,(0001) relies on two effect. First, a
binding hybride is formed, increasing the charge density at the N atom
(as is consistent with the positive change of work function upon
adsorption [31]). Second, the anti-bonding states are pushed above the
Fermi level and hence do not weaken the adsorbate-substrate bond.

  In conclusion, the atomic geometry of nitrogen adsorbed on
Ru\,(0001) has been explored by making use of LEED and DFT
calculation. Both techniques revealed that nitrogen atoms adsorb in
the threefold hcp sites on Ru\,(0001), the Ru--N bond length in the
($2\times 2$)-N and the ($\sqrt{3}\times\sqrt{3}$)R30$^\circ$-N phase
being about 1.93~\AA{}. The small Ru--N bond length is indicative of a
strong binding which is also evidenced by DFT calculations, indicating
binding energies of 5.82~eV and 5.59~eV for the ($2\times 2$)-N and
the ($\sqrt{3}\times\sqrt{3}$)R30$^\circ$-N phases, respectively. DFT
calculations show that a hypothetical (1$\times$1)-N phase on
Ru\,(0001) is not stable thermodynamically, while a hypothetical
(2x1)-N phase would be stable.

Acknowledgment:
We thank C. Stampfl and M. Scheffler for valuable discussions.  

\end{multicols}

\begin{table}
  \caption{Optimum Pendry r-factors for different structural models of
the Ru\,(0001)-N-($2\times 2$) and
Ru\,(0001)-N-($\sqrt{3}\times\sqrt{3}$)R30$^\circ$}
\begin{tabular}{ccc}
       & ($2\times 2$)-N & ($\sqrt{3}\times\sqrt{3}$)R30$^\circ$-N\\\tableline
on-top & 0.53 & 0.63 \\
bridge & 0.46 & 0.52 \\
fcc    & 0.70 & 0.59 \\
hcp    & 0.24 & 0.29 \\
\end{tabular}
\end{table}

\begin{table}
  \caption{Comparison of structural parameters obtained by LEED and DFT calculations. The expansion of the first Ru layer spacing is given with respect to the bulk layer distance.}
\begin{tabular}{c|cc|cc|}
method & \multicolumn{2}{c}{LEED} & \multicolumn{2}{c}{DFT calculations} \\
phase  & ($2\times 2$)-N & ($\sqrt{3}\times\sqrt{3}$)-N & ($2\times 2$)-N & ($\sqrt{3}\times\sqrt{3}$)-N \\ \tableline
Ru--N & ($1.93\pm 0.05$)~\AA{} & ($1.93\pm 0.06$)~\AA{} & 1.94~\AA{} & 1.94~\AA{} \\
Ru--N layer spacing & ($1.05\pm 0.05$)~\AA{} & ($1.10\pm 0.06$)~\AA{} & 1.08~\AA{} & 1.12~\AA{} \\
Ru expansion & ($-2\pm 2$)~\% & ($+1\pm 2$)~\% & 0~\% & $-0.5$~\% \\
adsorption site of N & hcp & hcp & hcp & hcp
\end{tabular}
\end{table}

\newpage

\begin{figure}
\psfig{figure=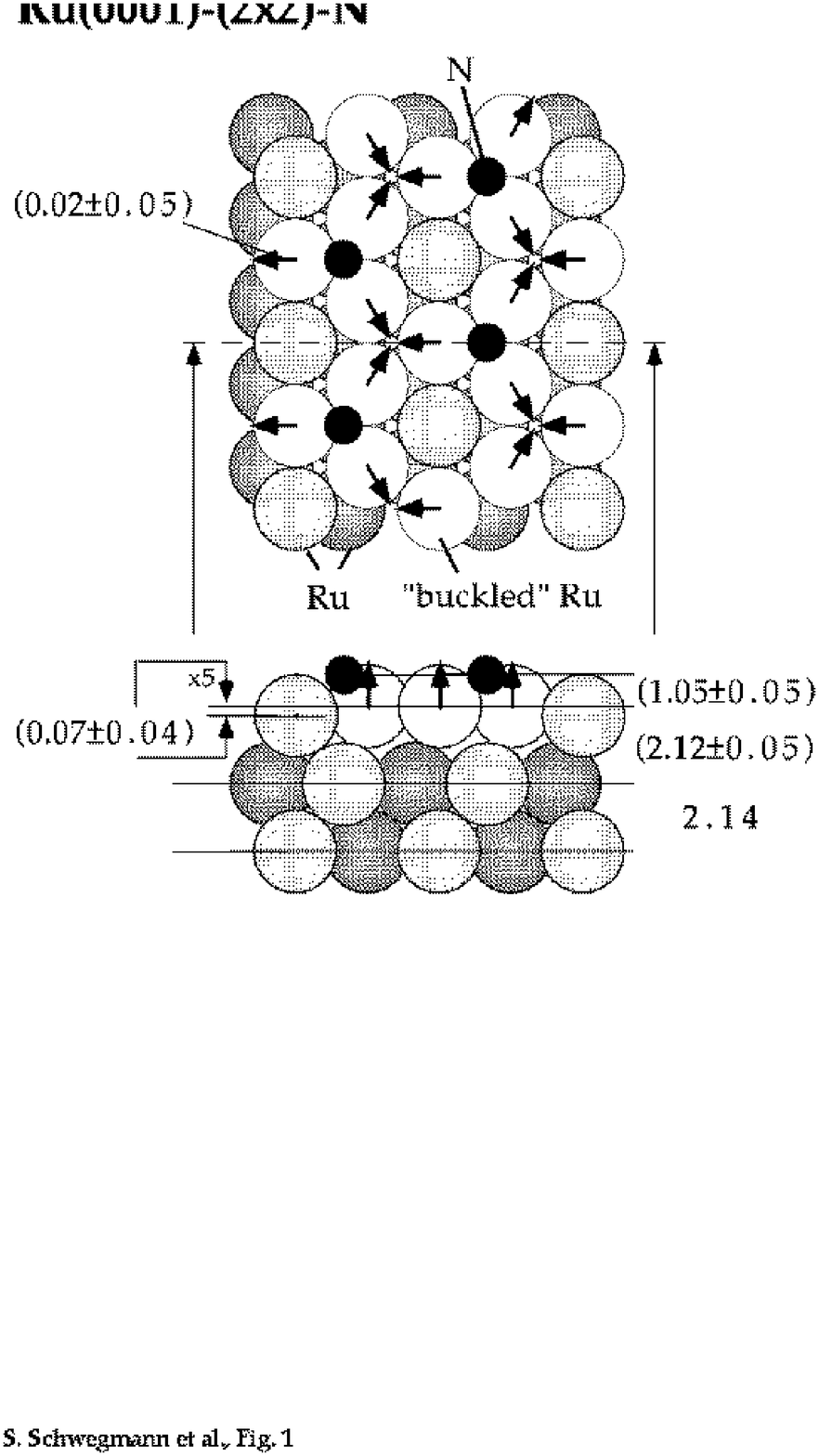}
  \caption{Hard-sphere model of the ($2\times 2$)-N structure on
Ru\,(0001). Atomic nitrogen resides in the threefold hcp hollow
position. The Ru--N bond length obtained by LEED is
1.93~$\pm$0.05~\AA{} while the value provided by DFT calculation is
1.94~\AA{}.}
\end{figure}

\clearpage

\begin{figure}
\psfig{figure=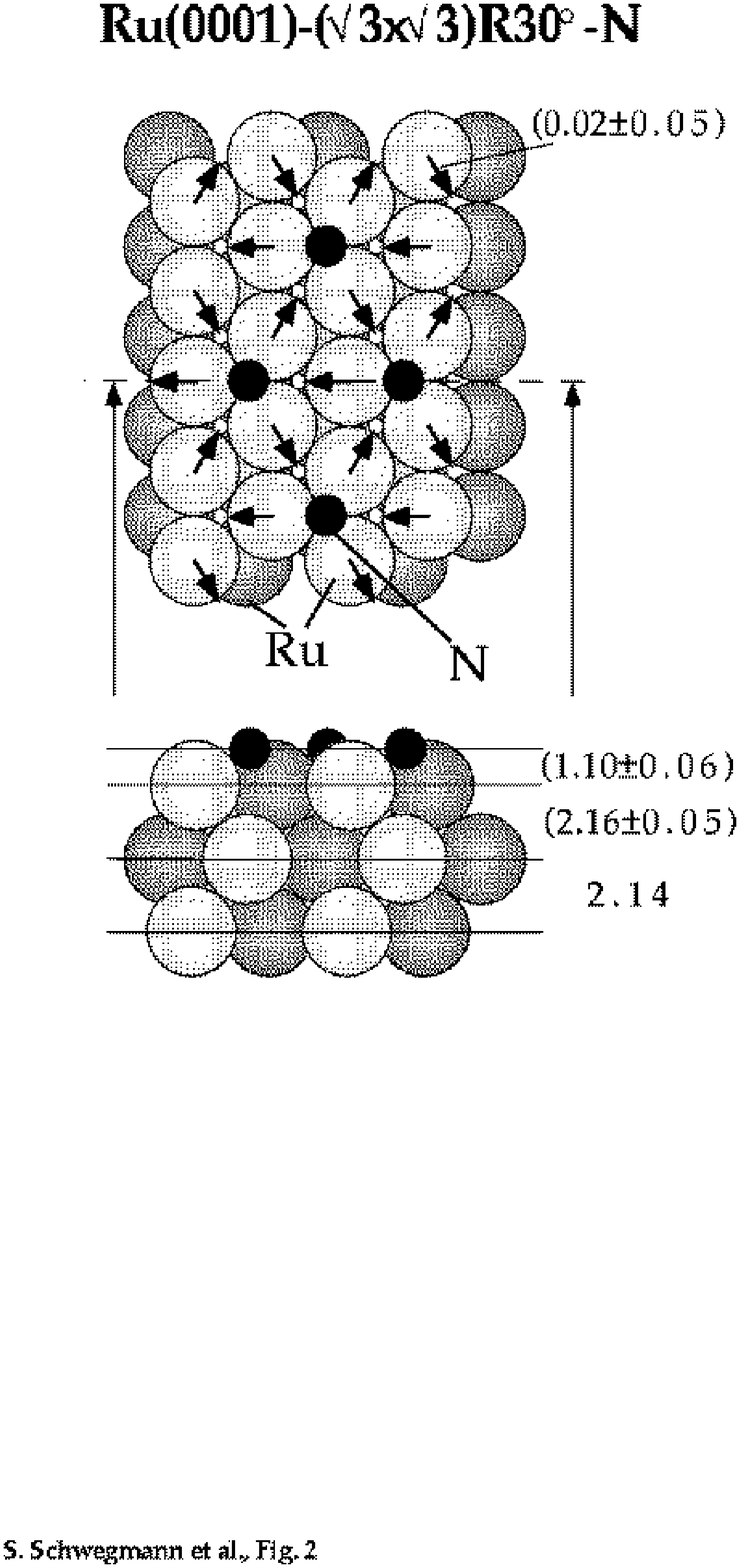}
  \caption{Hard-sphere model of the
($\sqrt{3}\times\sqrt{3}$)R30$^\circ$-N structure on Ru\,(0001).
Atomic nitrogen resides in the threefold hcp hollow site. The average
Ru--N bond length is found to be $1.93\pm 0.06$~\AA{} and 1.94~\AA{}
when using LEED and DFT calculations, respectively.}
\end{figure}

\clearpage

\begin{figure}
\psfig{figure=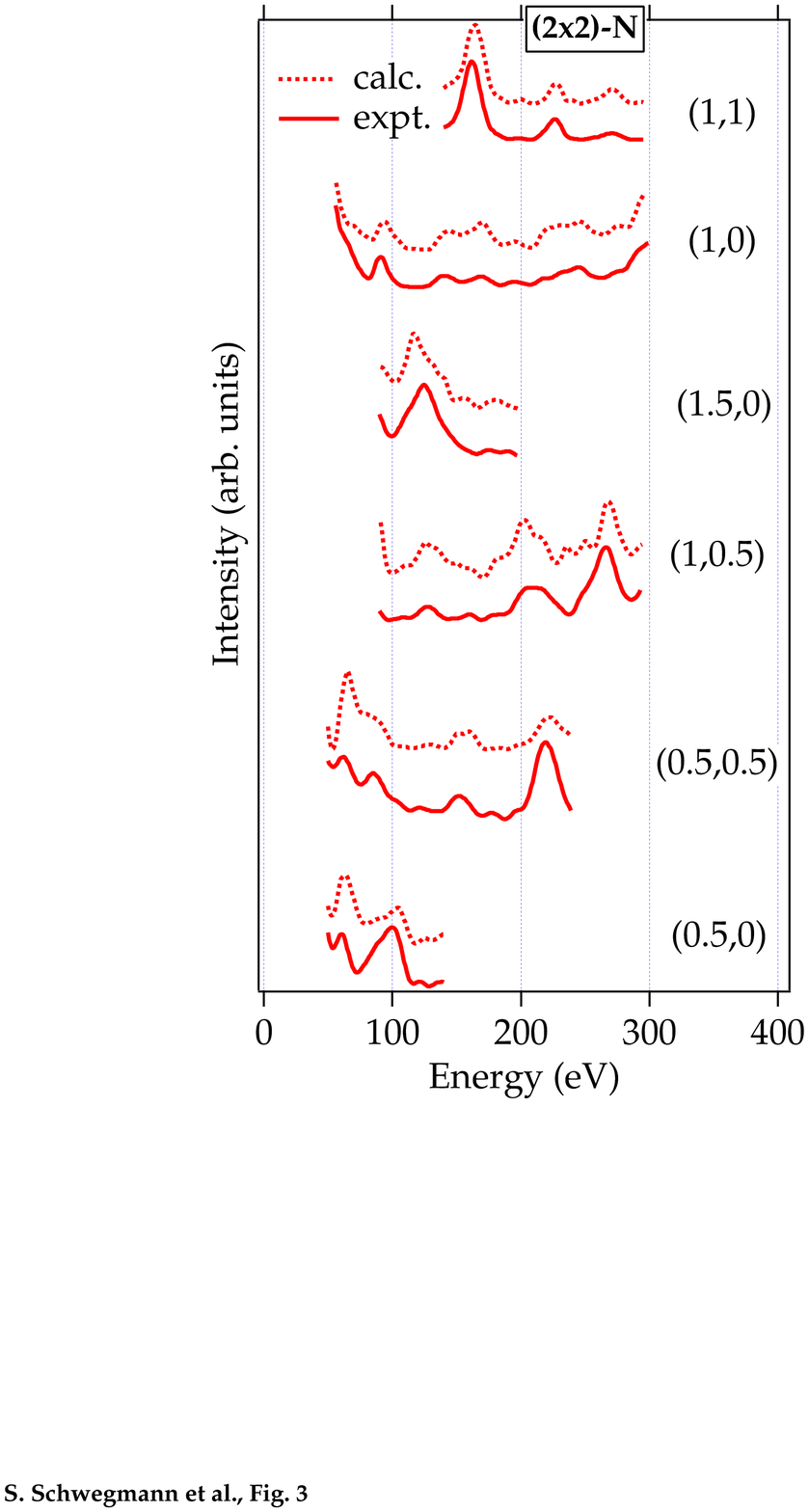}
  \caption{Comparison of experimental and calculated LEED~I(V) curves of
the Ru\,(0001)-($2\times 2$)-N structure. The overall Pendry r-factor is
R$_{\rm P}$~= 0.24.}
\end{figure}

\clearpage

\begin{figure}
\psfig{figure=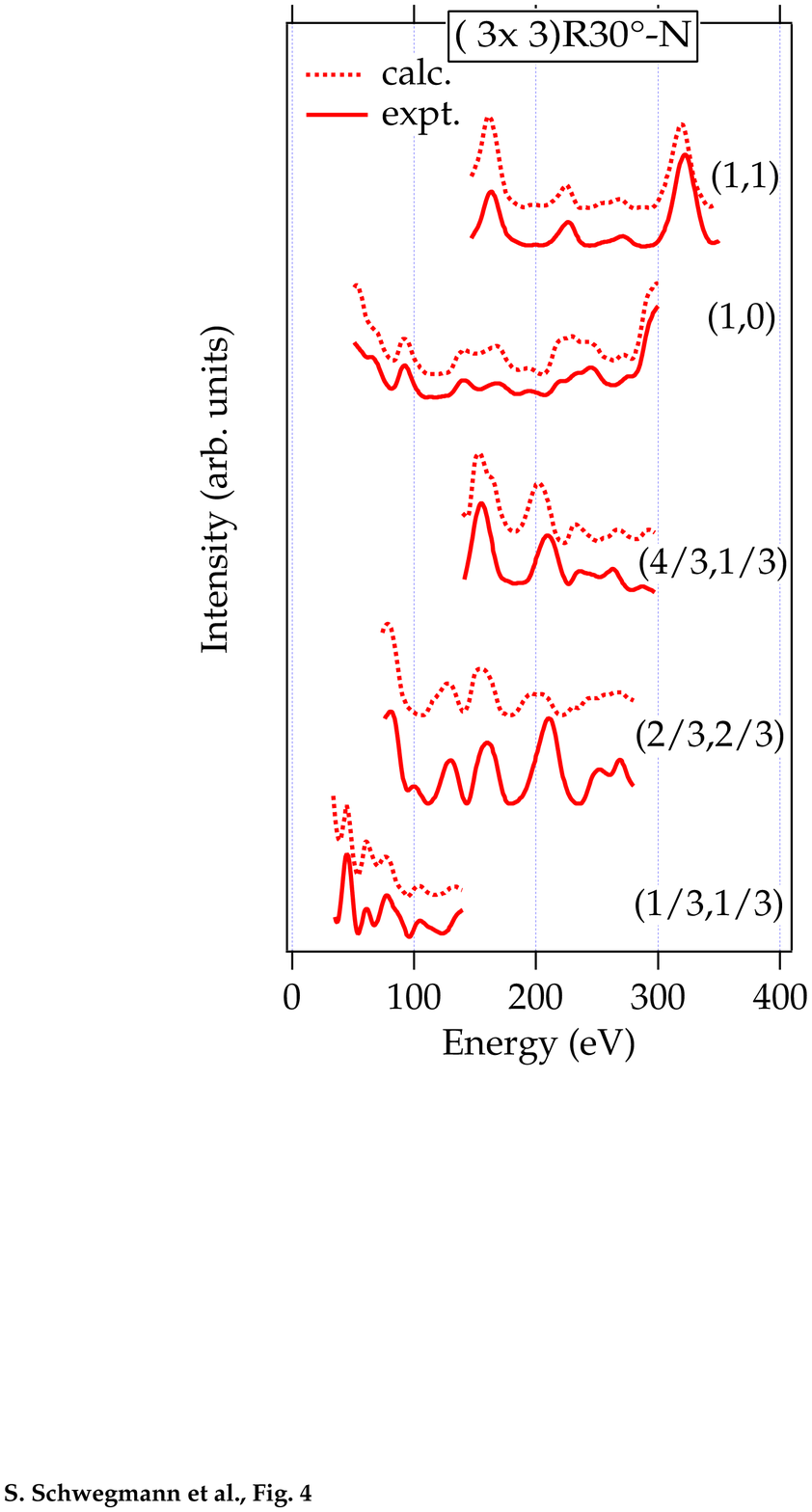}
  \caption{Comparison of experimental and calculated LEED I(V) curves of
the Ru\,(0001)-($\sqrt{3}\times \sqrt{3}$)R30$^\circ$-N structure. The
overall Pendry r-factor is R$_{\rm P}$~= 0.29.}
\end{figure}

\clearpage

\begin{figure}
\psfig{figure=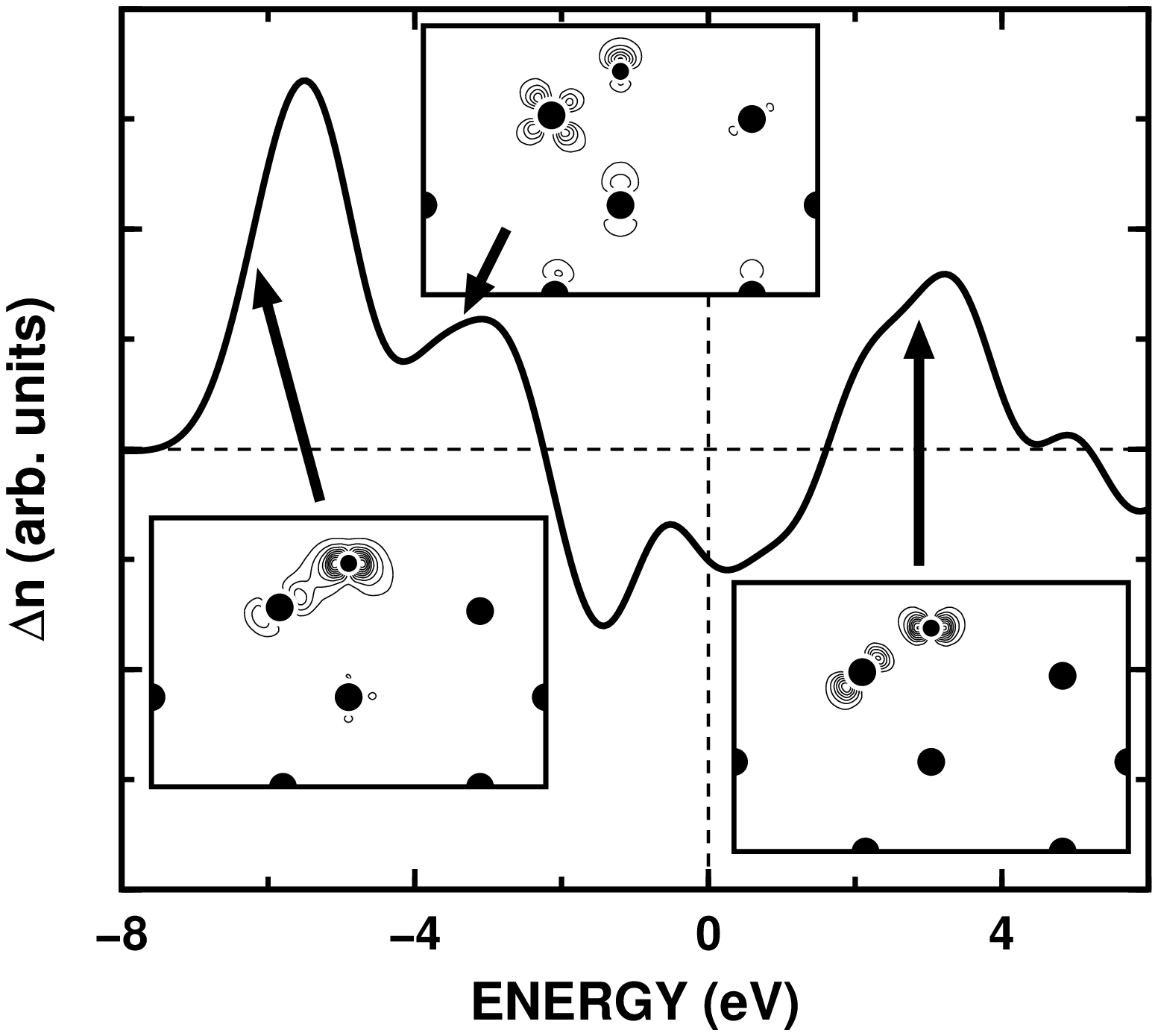}
  \caption{The difference in density of states, n(Ru+N)~--~n(Ru), where
n(Ru+N) is the density of states of the adsorbate system
Ru\,(0001)-($2\times 2$)-N and n(Ru) that of the clean Ru surface. The
one-electron eigenvalues are broadened by 0.6~eV. The density of a
representative bonding, non-bonding and anti-bonding orbital are shown
in the insets. The arrows point to the one-particle energy regions for
which the orbitals are shown. The positions of the N and Ru atoms are
indicated by small and large circles, respectively.}
\end{figure}

\end{document}